\documentstyle[12pt]{article}

\input epsf

\begin{document}

\def\frac#1/#2{\leavevmode\kern.1em
 \raise.5ex\hbox{\the\scriptfont0 #1}\kern-.1em
 /\kern-.15em\lower.25ex\hbox{\the\scriptfont0 #2}}
\def\degrees{\hbox{${}^\circ$\hskip-3pt .}}

\rightline{Fermilab-Pub-94/176-A}
\rightline{CfPA-94-TH-29}
\rightline{CWRU-P5-94}
\rightline{June 1994}
\baselineskip=16pt
\vspace{\bigskipamount}
\begin{center}
{\bf\large
Treating Solar Model Uncertainties: A Consistent Statistical
Analysis of Solar Neutrino Models and Data }
\end{center}
\vspace{\bigskipamount}

\begin{center}
Evalyn Gates\\
{\it Department of Astronomy \& Astrophysics and
 Enrico Fermi Institute\\
The University of Chicago, Chicago, IL 60637}\\

\vspace{\bigskipamount}
Lawrence M. Krauss\\
{\it Departments of Physics and Astronomy, Case Western Reserve University\\
Cleveland, OH 44106}
\vspace{\bigskipamount}

Martin White\\
{\it Center for Particle Astrophysics, University of California\\
Berkeley, CA 94720}
\end{center}

\vspace{\bigskipamount}

\centerline{{\bf Abstract}}
\vspace{\bigskipamount}
\noindent
We describe how to consistently incorporate solar model uncertainties,
along with experimental errors and correlations, when analyzing solar
neutrino data to derive confidence limits on parameter space for proposed
solutions of the solar neutrino problem.
Our work resolves ambiguities and inconsistencies in the previous literature.
As an application of our methods we calculate the masses and mixing angles
allowed by the current data for the proposed MSW solution using both Bayesian
and frequentist methods, allowing purely for solar model flux variations,
to compare with previous work.
We consider the effects of including metal diffusion in the solar models and
also discuss implications for future experiments.

\newpage

\noindent 1. {\bf Introduction}

As more experimental information has become available and theorists have
converged on ``new-physics'' explanations
\cite{White,BahcallBethe,Bah,BKL,ShiSchBah,KraPetMSW,KwoRos}
of the solar neutrino problem there has been interest in incorporating the
error budget of solar models into analyses of the data.
This has proceeded in stages.  The first incorporation of solar model flux
uncertainties did not take experimental correlations into account \cite{White}.
A subsequent analysis rectified this problem, but then did not properly account
for neutrino flux correlations \cite{BHKL}.  Most recently, a detailed analysis
has been performed which has largely resolved this latter problem by an
improved approximation for solar model uncertainties, a correct accounting for
experimental correlations, as well as a careful examination of such effects as
MSW \cite{MSW} mixing in the Earth in order to derive allowed regions of mass
and mixing angle \cite{HatLan2}.
Nevertheless, the general applicability of the approximations used there to
model solar model uncertainties is not guaranteed. In addition, the
determination of confidence limits and allowed regions of parameter space uses
a non-standard statistical analysis.

Now that it appears that the gallium results are stable and that no new
significant experimental light is likely to be shed on the problem until the
gallium experiments have been checked with neutrino sources (GALLEX is
scheduled for `calibration' in June 1994) or the next generation of detectors
comes on line in 3-4 years, there is time to consider a comprehensive,
consistent statistical analysis, vis a vis neutrino-based solutions of the
solar neutrino problem.
(Neutrino, rather than solar model, based solutions are now strongly indicated
by the present data, even without including the Homestake
results!\cite{ShiSchBah,KwoRos})
Such an analysis is the purpose of the present paper.
We shall demonstrate a technique which treats known solar model uncertainties
in a computationally simple  fashion, and then describe how to incorporate the
existing experimental information in order to derive confidence limits on
neutrino masses and mixing angles which have a well-defined statistical
meaning.  In the approximation in which all solar model uncertainties can be
parameterized in terms of the neutrino flux uncertainties, this technique
yields allowed regions in parameter space which can be compared with previous
results.

The determination of allowed regions requires four distinct parts:
(1) a calculation of solar model uncertainties,
(2) a model of neutrino transport and detection probabilities,
(3) a determination of experimental uncertainties and correlations,
and finally
(4) a well defined statistical procedure for comparing predictions and
observations.

The outline of the paper is as follows.  We first describe solar model
uncertainties gleaned from Monte Carlo studies of solar models.
We demonstrate that the essential information about this type of solar model
uncertainty
is contained in the neutrino flux correlation matrix, which can be calculated
either directly using the solar models themselves, or else using a simple but
well defined approximation.
Next we demonstrate how to translate these flux correlations into an
experimental covariance matrix necessary to properly incorporate the
experimental error budget.  Following this we describe, for both MSW and
vacuum oscillations, how one derives survival probabilities following transport
through the sun and earth.  Finally, we describe how to consistently derive
allowed regions for neutrino parameter space using well defined statistical
probes in a way which avoids problems with past analyses, and discuss the
meaning of our results for future experiments and particle physics models.

It is worth emphasizing in advance that by outlining a well defined statistical
procedure for comparing theory and observation we do not necessarily subscribe
to the view that only statistical solar model uncertainties are relevant, or
even that they may be the most important uncertainties.  It is quite
possible, indeed perhaps likely, that systematic solar model uncertainties, due
primarily to the introduction of new physics into the model (such as heavy
metal
diffusion---see below) could shift the {\it entire} allowed range of model
parameters determined by the methods we describe here.
Nevertheless, as the standard solar model gets more complete, this will be less
likely.  Our purpose here is to define a consistent and correct procedure
which may be applied as both the data and the theoretical models improve.

\vspace{\bigskipamount}

\noindent 2. {\bf Solar Model Uncertainties}

Comprehensive estimates of the present solar model uncertainties have been made
by Bahcall and collaborators \cite{BU,TheBook}, who performed detailed Monte
Carlo analyses of the neutrino fluxes that result when solar model input
parameters are varied over their allowed ranges.
Since calculating many full solar models can prove cumbersome in terms of
computing time, it is useful to have a reliable and efficient approximation
scheme which reproduces the results of such a calculation.
Several schemes have been proposed which account for the variation in the
total flux of neutrinos, which is in general the major source of uncertainty
(from solar physics) in the prediction of the experimental rates.

Two different approaches have been applied to this problem.
The first involves simplifying the solar model parameter space, an example of
which we will call ``the $T_c$ approach'' \cite{BHKL}.  Here the fluxes are
parameterized by a single (solar model {\em output}) variable -- the core
temperature of the sun: $T_c$. The temperature dependence of the various fluxes
are derived from the scatter plots of flux vs $T_c$ from solar model Monte
Carlo calculations. (See e.g.~figures 6.2 and 6.3 in ref \cite{TheBook}.)
Approximating the temperature dependence of the fluxes by power laws in $T_c$
specifies the flux distributions, with the error in $T_c$ determined so as to
give the appropriate uncertainties in the fluxes.

While the scatter plots indicate that the relationship between the neutrino
flux and $T_c$ can be approximately described by a simple power law, this
relationship is only approximate and there remains a significant width to the
straight line that would describe a perfect power law dependence.
Because of this width, these plots do not indicate how the various fluxes are
correlated.  For example, a solar model with a higher $T_c$ may correspond to
an increased $pp$ flux, but little or no corresponding decrease in the ${}^8$B
flux.  The $T_c$ method, based on only {\em one} parameter, of course produces
totally (anti-)correlated uncertainties for the neutrino fluxes while the solar
model flux uncertainties exhibit a wide range of correlations.
The differences in the correlations for the $T_c$ parameterization and the full
solar model Monte Carlo are a reflection of the scatter in the plots of
\cite{TheBook}.  By overestimating the correlations, the $T_c$ approach tends
to underestimate the size of the allowed parameter region for a given
confidence level. Thus while a $T_c$ parameterization can be a useful tool
in some instances, it is not appropriate for calculating solar model
uncertainties \cite{BahcallBethe}.

An updated version of this method \cite{HatLan2} includes not only $T_c$ but
also cross section uncertainties in the form of two extra parameters, chosen
from among the (nuclear cross section) input parameters to the solar models.
This allows this method to be tuned to more closely approximate
the full solar model correlations \cite{HatLan2}.
Any method which reproduces the flux correlation matrix can correctly include
the solar model errors, so this updated method and the ``power law method''
(described below) should agree on the statistical content of the solar
model uncertainty.
However, the applicability of this method, including the determination of
which combination works, and which $T_c$ uncertainty to use can only strictly
be determined after the fact by explicitly utilizing the detailed results of
the full solar model Monte Carlo calculations.

An alternative approach, proposed earlier \cite{White}, parameterizes the
solar model uncertainties in terms of the logarithmic derivatives of the
fluxes with respect to the 10 solar model {\em input} parameters.
It was shown in \cite{TheBook} that for small variations in the input
parameters the neutrino fluxes, $\phi$, can be expressed as
\begin{equation}
  \phi_j \propto \prod_k \left(\Gamma_k\right)^{\alpha_{j,k}}
\label{eqn:fluxreln}
\end{equation}
where $\alpha_{j,k}$ is the logarithmic partial derivative of $\phi_j$
($j=pp$, pep, hep, ${}^7$Be, ${}^8$B, ${}^{13}$N, ${}^{15}$O and ${}^{17}$F)
with respect to the input parameter $\Gamma_k$.  The solar model flux
uncertainties can thereafter be obtained from a Monte Carlo procedure assuming
Gaussian distributions for the input parameters, as described in more detail in
\cite{White}.  This method has a firm basis in describing the errors in the
output function (solar model fluxes) in terms of the errors in the input
parameters, and the $\alpha_{j,k}$ are readily available \cite{TheBook}.
We shall refer to this as ``the power law approach''.

{}From a Monte Carlo analysis using this approach, we obtain an estimate of
the theoretical uncertainties in the predicted fluxes for each species.
This allows us to determine the correlations between the various fluxes, which
will be important for computing correlations between the rates predicted for
different detectors.
The elements of the covariance matrix for the various fluxes ($\phi_j$) are
given by \cite{statbook}
\begin{equation}
  V_{jk} = \left\langle(\phi_j-\bar\phi_j)(\phi_k-\bar\phi_k)\right\rangle
\label{eqn:covariance}
\end{equation}
where the angled brackets indicate an average over the solar models and
$\bar\phi=\left\langle\phi\right\rangle$.
To display the correlations we present the correlation matrix, whose elements
are given by
\begin{equation}
  \rho_{jk} = {V_{jk}\over \sigma_j \sigma_k} ,
\end{equation}
where $\sigma_j=\sqrt{V_{jj}}$ is the standard deviation in $\phi_j$.  Note
that in the correlation matrix the diagonal elements are 1 by definition and
off-diagonal elements are equal to $(-)1$ in the limit of perfect
(anti-)correlation.

It is important to note that the partial derivatives in (\ref{eqn:fluxreln})
were determined via the solar model with fixed solar luminosity.
Even though the power law approach does not explicitly enforce such a
constraint, the use of the relation (\ref{eqn:fluxreln}) will result in a
covariance matrix equal to that from the fully self-consistent solar model
Monte Carlo calculations.
This was implicitly exploited in our previous work \cite{White,KGW} and is
explicitly demonstrated in tables \ref{tab:bu-correlation} and
\ref{tab:mc-correlation} which compare the covariance matrices for the two
approaches.
The agreement between our Monte Carlo calculation and the $1{,}000$ models of
Bahcall \& Ulrich is good, as one would expect, except for the hep neutrinos.
Since the hep and ${}^{17}$F contribute negligibly to the rate in all the
detectors this is not of concern.
The flux correlation matrix for the $T_c$ approach has all elements equal to
$\pm1$ since there is perfect correlation---i.e. all the fluxes depend on one
parameter.  This is relaxed in the updated approach including the cross
section uncertainties \cite{HatLan2} which it is claimed also reproduces
table \ref{tab:bu-correlation}.

As we shall discuss later, at least as far as flux uncertainties are concerned,
the entire statistical content of the full solar model Monte Carlo calculation
is contained in the covariance matrix $V_{ij}$.  Our method is designed to
reproduce this matrix based on the matrix of flux derivatives, while the
updated $T_c$ approach reproduces this matrix by a posteriori construction.
Nevertheless, once the matrix $V_{ij}$ is obtained from the solar models, this
alone is sufficient, and there is no need for either approximation.
For this reason, this quantity is as important to extract from solar model
calculations as are logarithmic flux derivatives $\alpha_{j,k}$, and we
suggest that future work on solar model calculations include the results
for $V_{ij}$ explicitly.

In our fits, to be described later, we use the updated fluxes from the
Bahcall \& Pinsonneault solar model \cite{BP} which incorporates Helium
diffusion and new equation of state and opacity calculations.
Although a full Monte Carlo treatment of the flux uncertainties has not been
performed for this model we have updated the correlation matrix
shown in table \ref{tab:mc-correlation} to incorporate the errors on the input
parameters as given in \cite{BP}.  This does not include the uncertainty in
the fluxes from variations in diffusion, but is the best use of currently
available information.

\vspace{\bigskipamount}

\noindent 3. {\bf Experimental Rate Uncertainties}

The central quantity to use in determining how well model predictions agree
with the observed rates will be the rate covariance matrix.  When solar model
uncertainties can be completely parameterized in terms of the flux covariance
matrix,  the covariance matrix for the rates can be calculated directly from
that for the fluxes.
In this case, for any theoretical model the predicted rate in the detector
$R_a$ ($a=$H,K,Ga) is a linear combination of the fluxes $\phi_j$ with
coefficients functions of the theory parameters (see equation
(\ref{eqn:rate})).
If we write $R_a=r_{aj}\phi_j$ with
$r_{aj}=r_{aj}(\Delta m^2,\sin^2 2\theta)$ then it is straightforward to show
that
\begin{equation}
  V_{ab} = \sum_{jk} r_{aj} r_{bk} V_{jk}
\label{eqn:rcovariance}
\end{equation}
It is important to note that at this stage experimental uncertainties,
including those from detection cross section uncertainties have not been
introduced.

The correlation matrices for both the fluxes and the experiments (assuming
standard model interactions for the neutrinos) are shown in tables
\ref{tab:bu-correlation} and \ref{tab:mc-correlation} for the Bahcall \& Ulrich
standard solar model(s) \cite{BU} and our power-law approach.
Note that the experimental rates are almost perfectly correlated (a fact which
was ignored in our earlier work \cite{White}). The correlation between
experiments can decrease once neutrino mixing is allowed.  For example the
correlations can be as low as $\sim25\%$ for
$(\Delta m^2,\sin^2 2\theta)$ in the small-angle allowed region (see
figure~\ref{fig:allowed}).  However generally the correlation is above 80\%.
It is initially surprising that the rates for Homestake and Kamiokande, which
measure principally ${}^8$B neutrinos, should be (strongly) {\em positively}
correlated with Gallium, which measures principally $pp$ neutrinos, when
${}^8$B and $pp$ neutrinos are strongly {\em anti}-correlated!
The resolution of this apparent paradox is that while the major contribution
to the Gallium {\em rate} is due to $pp$ neutrinos, the ${}^8$B and ${}^7$Be
neutrino fluxes are much more uncertain and are the principal contribution to
the {\em uncertainty} in the Gallium rate.  For Gallium
\begin{equation}
{\rm R_{Ga}} = 71\ pp + 34\ {}^7{\rm Be} + 14\ {}^8{\rm B} + \cdots
\ {\rm SNU}
\end{equation}
The relative errors of the $pp$, ${}^7$Be and ${}^8$B fluxes are $\frac 1/2\%$,
$5\%$ and $15\%$ respectively.  Clearly the uncertainty in ${}^7$Be and
${}^8$B dominates the uncertainty in the Gallium rate.

\vspace{\bigskipamount}

\noindent 4. {\bf Neutrino Transport}

In order to determine the experimental rate matrix described above, we must
utilize analytic or numerical techniques to propagate neutrinos through the
sun, empty space, and the earth in order to determine survival probabilities
and resulting flux modulations.  The methods used differ, depending upon
whether one is interested in the region of mass and mixing angle space where
MSW oscillations or vacuum oscillations are important.

\vspace{\bigskipamount}

\noindent {\bf a. Vacuum oscillations}

In addition to the MSW model, there exists the possibility that the observed
deficit of neutrinos could be due to oscillations in vacua between the
sun and the earth \cite{vacuum}.
In this case the details of the production in the sun are unimportant and we
need keep track only of total flux variations.
The survival probability is \cite{TheBook}:
\begin{equation}
  P(\nu_e \rightarrow \nu_e) =
  1-\sin^2 2\theta \sin^2{\pi L\over L_V}
\end{equation}
with $L_V=4\pi E/\Delta m^2$.  Additionally one can average this
survival probability over the change in the Earth-Sun distance during
times comparable with the average duration of an experimental ``run''.
We find our conclusions do not depend on the averaging.

Performing a fit to the current experimental data (as described later)
we find a small region in parameter space which is allowed at the 95\%
confidence level.  This region agrees in general with those found by other
authors \cite{Hata,KPvac}, who have explored this theory in detail, and
we will have nothing further to say about it.

\vspace{\bigskipamount}

\noindent {\bf b. MSW oscillations}

Perhaps the most promising neutrino mixing solution to the solar neutrino
problem is the Mikheyev-Smirnov-Wolfenstein (MSW), or matter-enhanced mixing,
model \cite{MSW,TheBook}.
In this section we give details of our approximations and modelling of this
effect in relation to computing the predicted rates in the Homestake
\cite{Homestake}, Kamiokande \cite{Kamiokande1} and Gallium \cite{Sage,Gallex}
neutrino experiments.

To compute the rate predicted by a model for any detector we need information
about the neutrino production in the sun.  We use the flux distributions over
the production regions $d\phi_i(r)$ and the electron number density as a
function of solar radius, $N_e(r)$, from \cite{TheBook} and the scale heights
at resonance $r_0$ tabulated in \cite{KP} for use with their analytic
approximations\footnote{We correct a programming error in our earlier work in
which $r_0$ was incorrectly read from the table.}.
We have explicitly checked that using
$r_0\equiv N_e^{\rm res}/|dN_e/dr|_{\rm res}$ from the Bahcall \& Pinsonneault
standard solar model \cite{BP} produces the same results.
Additionally we assume that the energy spectrum of neutrinos at each $r$ is as
described in \cite{TheBook}. We have fitted the spectra for all species as a
polynomial times the relevant $\beta$-decay spectrum (correcting the
typographical error in eq.~8.15 of \cite{TheBook}). These values are then input
into the analytic expressions for the $\nu_e$ survival probability \cite{KP}
(see also \cite{probs}), as outlined in our previous work \cite{KGW} and
summarized below.

If the neutrino passes through a resonance on its way through the sun then we
define
\begin{equation}
  4n_0 = r_0 \left( {\Delta m^2 \over {2E}} \right)
  \left( {\sin^2 2\theta \over\cos 2\theta} \right),
\label{eqn:n0}
\end{equation}
where $N_e(r)$ is the electron density profile in the sun.
The electron density at resonance is given by
\begin{equation}
 N_e^{\rm res} = \left( {\Delta m^2 \over {2E}} \right)
 \left( {\cos 2\theta \over\sqrt{2}G_F} \right).
\label{eqn:Nres}
\end{equation}
In terms of $n_0$ we classify the transition as either adiabatic
($4n_0\gg 1$) or non-adiabatic ($4n_0\leq 1$).

Let $N_e^{(0)}$ be the electron density at the point of $\nu$ production,
then for neutrinos in the adiabatic region, or those in the non-adiabatic
region with $N_e^{\rm res}< N_e^{(0)}/(1 + \tan 2\theta)$ the analytic
expression for the $\nu_e$ survival probability is given by
\begin{equation}
  P(\nu_e\rightarrow\nu_e) =
  {1\over2} + \left( { 1 + e^{-x} \over 1 - e^{-x} } \right)
  \left[ {1\over 2} - {e^{-y}\over 1 + e^{-x}}\right]
  \cos 2\theta_m \cos 2\theta,
\label{eqn:PNA}
\end{equation}
where
\begin{eqnarray}
             x & = & 2\pi\ r_0{\Delta m^2 \over 2E} \\
             y & = & 2\pi\ n_0(1 - \tan^2 \theta) \\
\cos 2\theta_m & = & (1 - \eta) / \sqrt{ (1-\eta)^2+\tan^2 2\theta}\\
          \eta & \equiv & N_e^{(0)} / N_e^{\rm res}.
\end{eqnarray}

For neutrinos in the non-adiabatic region produced near resonance,
$N_e^{(0)}/(1+\tan2\theta)\leq N_e^{\rm res}\leq N_e^{(0)}/(1-\tan2\theta)$,
the corresponding expression for the survival probability is
\begin{equation}
  P(\nu_e \rightarrow \nu_e) =
  {1\over 2} \left[ 1 + \exp (-\pi n_0)\right],
\label{eqn:P0NA}
\end{equation}
while for non-adiabatic transitions with
$N_e^{\rm res}>N_e^{(0)}/(1-\tan2\theta)$ or adiabatic transitions with
$N_e^{\rm res}>N_e^{(0)}$ we use
\begin{equation}
  P(\nu_e \rightarrow \nu_e) =
  {1\over 2} + {1\over 2} \cos2\theta_m \cos2\theta.
\end{equation}
We have also included in our analysis the effects of double resonances in
the sun, see \cite{KGW}.

\vspace{\bigskipamount}

\noindent {\bf c. Earth Effects}

It has long been known \cite{earth} that for $\Delta m^2$ near
$10^{-6}{\rm eV}^2$ and large $\sin^22\theta$ it is possible to `regenerate'
$\nu_e$ by having neutrinos pass through the earth.
The survival probability $P(\nu_e\rightarrow\nu_e)$ is very sensitive to the
path length of the neutrinos in the earth and so neutrinos with parameters in
this range should give rise to day/night and seasonal variations in the
observed flux.
Since no such effect has been seen \cite{Kamiokande1} this serves to rule
out a region of parameter space near $\Delta m^2\sim10^{-6}{\rm eV}^2$
and $\sin^22\theta\sim 0.2$.

We follow \cite{HatLan1} in including this ``Earth effect'' in our fits,
though our treatment differs from theirs.  Rather than keep track of the
predicted dependence of $P(\nu_e\rightarrow\nu_e)$ on the path length
(which changes during the ``night'') and fit to the data in many
bins\footnote{We note in passing that the binned data of \cite{Kamiokande1}
for the day/night effect has a very low $\chi^2$ per degree of freedom,
which may indicate correlated (systematic) uncertainties in this data set.
In any case such a low $\chi^2$ will bias a fit in which these points form
most of the degrees of freedom.},
we choose to use a number which summarizes that no effect is actually
seen.  Consequently we use the quoted measurement of \cite{Kamiokande1}
\begin{equation}
\left.
\left\langle {{\rm day}-{\rm night}\over{\rm day}+{\rm night}}\right\rangle
\right|_{\displaystyle {\rm year}} = -0.08\pm 0.11
\label{eqn:daynightnumber}
\end{equation}
which is independent of the solar model flux uncertainties.  Since this
quantity does not depend on the neutrino flux we can simply add the $\chi^2$
from this fit to the $\chi^2$ obtained from fitting to the time average
rates as will be described later.  The effect will be to rule out a region
of parameter space where a large day/night effect would be predicted.

To predict the l.h.s.~of (\ref{eqn:daynightnumber}) we follow
\cite{HatLan1,BalWen}.  Since only the integrated electron density along the
line of sight matters for the average $P(\nu_e\rightarrow\nu_e)$ we model
the Earth as 5 concentric shells of constant electron density $N_e$,
which we have taken from the models of \cite{earthmodel} and listed in
table \ref{tab:earthmodel}.  Including the Earth effect the survival
probability of a $\nu_e$ which has MSW survival probability $P_{\rm MSW}$
is given by \cite{HatLan1,BalWen}
\begin{equation}
 P_E = P_{\rm MSW} |a|^2 + (1-P_{\rm MSW})|b|^2 +
  ({1\over2}-P_{\rm MSW})\tan2\theta(ab^{*}+ba^{*})
\end{equation}
where $a$ and $b$ are elements of the unitary matrix which describes the
evolution of neutrinos through the earth
\begin{equation}
\left( \begin{array}{c} \nu_e(t) \\ \nu_x(t) \end{array} \right)
= \left( \begin{array}{cc} a & b \\ -b^{*} & a^{*} \end{array} \right)
\left( \begin{array}{c} \nu_e(0) \\ \nu_x(0) \end{array} \right)
\end{equation}
Once we have solved for this matrix for an arbitrary shell of our model Earth
we can obtain the full matrix by multiplication of the evolution matrices for
each shell in the appropriate order (entering and leaving).  Thus the
problem reduces to calculating $a$ and $b$ for propagation through a shell
of constant $N_e$.  Dropping a constant energy offset from $\nu_e$ and $\nu_x$,
which contributes only an irrelevant overall phase, the evolution equation is
\begin{equation}
i{d\over dt}
\left( \begin{array}{c} \nu_e \\ \nu_x \end{array} \right) =
\left[
\left( {G_F N_e\over\sqrt{2}}-{\Delta m^2\over 4E}\cos2\theta\right)\sigma_3
+{\Delta m^2\over 4E}\sin2\theta\ \sigma_1
\right]
\left( \begin{array}{c} \nu_e \\ \nu_x \end{array} \right)
\end{equation}
where $\sigma_i$ are the Pauli matrices.  We solve this using the identity
$\exp[i\vec{a}\cdot\vec{\sigma}]=\cos|a|{\bf 1}+
i\sin|a|(\hat{a}\cdot\vec{\sigma})$
to yield
\begin{equation}
\begin{array}{ccc}
a & = & \cos|h| - i\hat{h}_3\sin|h| \\
b & = & -i\hat{h}_1\sin|h|
\end{array}
\end{equation}
with
\begin{equation}
\vec{h}=\left( {\Delta m^2\over 4E}\sin2\theta,\ 0,
\ {G_F N_e\over\sqrt{2}}-{\Delta m^2\over 4E}\cos2\theta\right)\times
\ {\rm path\ length}
\end{equation}
and $\hat{h}=\vec{h}/|h|$.  Although our model is relatively crude, given that
we are trying to fit to the
absence of an effect it is sufficient for our purposes.

The final task is then to integrate over the paths through the Earth during
the course of the night/year.  In our model Earth with spherical symmetry
the path is totally defined by giving the angle $\theta_0$ subtended at the
center of the Earth by the point of entry of the $\nu$ beam and the detector.
In the limit that the Earth-Sun distance is much larger than the Earth's
radius we have that
\begin{equation}
\cos\left( {\pi - \theta_0\over 2} \right) =
\sin\delta\sin i + \cos\delta\cos i\cos(\phi+\pi)
\label{eqn:day-night}
\end{equation}
where $\phi$ is the azimuthal angle between the Sun and the detector as
measured from the center of the Earth, $\delta$ is the detector latitude
and $i=23\degrees5\sin(\omega_{\odot}t)$ is the inclination of the ecliptic to
the Earth's equator.
Averaging over $\phi$ and $\omega_{\odot}t$ we obtain the distributions for
$\theta_0$, with which we can then determine the $\nu_e$ survival probability
averaged over night/year.  For a given mass and mixing angle, we compare this
value with the r.h.s.~of (\ref{eqn:daynightnumber}) in determining the $\chi^2$
fit to the data.
The survival probability $P(\nu_e\rightarrow\nu_e)$ including the Earth
effect is shown in figure \ref{fig:regeneration} for a parameter set which
can be compared with Fig.3b of \cite{BalWen}.

\vspace{\bigskipamount}

\noindent {\bf d. Calculating the rates}

Using the above and the formulae for $P_{\rm MSW}$ outlined in the previous
section we computed the survival probability averaged over the night and the
year.  These probabilities and the fluxes for each species, $j$, are then
convolved with the detector response $D_i(E_\nu)$
\cite{White,TheBook,KGW,Gates,Hirata}
for neutrinos of flavor $i$ and energy $E_\nu$ to get the predicted rate
\begin{equation}
  R = \sum_{ij} \int dE_\nu\ \phi_j(E_\nu) P_i(E_\nu) D_i(E_\nu)
\label{eqn:rate}
\end{equation}
for each model.
We include the contributions from $j=pp$(10), pep(1), ${}^7$Be(2),
${}^8$B(30), ${}^{13}$N(20) and ${}^{15}$O(20) neutrinos, where the number
in parenthesis after each species is the number of energies computed for
each spectrum in the integration.  The contribution from hep and ${}^{17}$F
neutrinos are less than ${\frac 1/2}\%$ for all the experiments and can be
safely ignored.
Our results for the iso-SNU contours for the Homestake, Kamiokande
and Gallium experiments compare well with those in \cite{HatLan2}.

\vspace{\bigskipamount}

\noindent 5. {\bf Data \& Model Testing}

We use the latest data for the time-averaged rate in the Homestake
\cite{Homestake}, Kamiokande \cite{Kamiokande1}, SAGE \cite{Sage} and
GALLEX \cite{Gallex} experiments.
Since the theory predictions for the SAGE and GALLEX experiments are identical
and the experimental values agree within errors, we have combined the two rates
($74\pm20$ and $79\pm12$ SNU for SAGE and GALLEX respectively) in our fit.
We have added the statistical and systematic errors in quadrature, since they
are independent.  The assumption of a gaussian distribution for the systematic
error is problematic because, by its very nature, the systematic error has no
statistical distribution.  However, if we regard the gaussian as representing
the state of our knowledge about the systematic error then clearly a function
which penalizes large `errors' is more appropriate than a flat distribution
(which would correspond to maximal ignorance of the size of the systematic
error).  We have chosen a gaussian for simplicity.
The three experimental rates, as a fraction of the standard solar model
predictions, are shown in table 2. 
In our analysis we have added the cross section  uncertainties \cite{BU} in
quadrature to the quoted errors.
For Gallium this ignores the energy dependence of the uncertainty from the
resonance, but this uncertainty affects primarily the ${}^8$B contribution
to the rate which is already small and which we expect to be suppressed for
the masses and mixing angles of interest to us.

We have used two parametric methods: a $\chi^2$ goodness-of-fit procedure and
a Bayesian likelihood analysis \cite{statbook} to compare the measured rates
$R_a^m\pm\sigma_a$ ($a$=H,K,Ga) to the rates predicted by the model
($\Delta m^2,\sin^2 2\theta$).
Both methods rely on the assumption that the errors in the solar model
predictions are gaussian under small variations in the solar model input
parameters, which appears to be a good assumption \cite{TheBook}.
For more discussion of the methods we use see
\cite{statbook,statart,Berger,Edwards}.

Including the solar model uncertainties, each set of MSW parameters
$(\Delta m^2,\sin^2 2\theta)$ defines a distribution of rate triplets $R_a$.
One can calculate the covariance matrix $V_{ab}^{SM}$ for the triplets
analogously to equations (\ref{eqn:covariance},\ref{eqn:rcovariance}).
To the theoretical solar model covariance, $V_{ab}^{SM}$, we add the
experimental errors, $\sigma_a$, to obtain the full covariance matrix:
$V_{ab} = V_{ab}^{SM} + \sigma_a^2\delta_{ab}$.
Defining \cite{statbook}
\begin{equation}
    \chi^2 \equiv (R_a-\overline{R}_a) V_{ab}^{-1} (R_b-\overline{R}_b) ,
\label{eqn:chi2def}
\end{equation}
where $\overline{R}$ is the standard solar + neutrino-mixing model rate
prediction, the distribution of $\chi^2$s defined by the theory is a chi-square
distribution with 3 degrees of freedom.
The statement that the theory is ruled out at some confidence level is the
claim that the $\chi^2$ for the measured triplet lies in the large-$\chi^2$
tail of the distribution defined by the theory, i.e.~that the measured value
is unlikely.
If the {\em data} is within the 95\% confidence level of a given theory we say
the {\em theory} parameters are allowed at the 95\% confidence level by the
data.

Previous authors have generally implemented instead a {\em best-fit}
procedure that attempts to estimate the parameters $\Delta m^2$ and
$\sin^2 2\theta$ from the data and assign errors to the inferred values.
One takes the parameters which minimize $\chi^2$ as the central values,
with an {\em allowed range} given by the condition that
$\chi^2(\Delta m^2,\sin^2 2\theta)<\chi^2_{\rm min}+\nu$,
with $\nu$ determined by the range of $\sigma$ desired and the number of
parameters being estimated.
Such an approach is based on the maximum likelihood procedure under the
assumption that the correlation matrix is independent of the parameters being
estimated.
(This assumption is obviously not true for this case, but the errors
introduced turn out to be numerically small.)
This approach makes the additional assumption that
$\chi^2(\Delta m^2,\sin^2 2\theta)$ is well approximated by a quadratic over
the relevant range of parameters.
As can be seen in figure \ref{fig:allowed} this assumption is clearly false
over the range of $(\Delta m^2,\sin^2 2\theta)$ of interest.
It is important to realize that the statistical answers one gets depend upon
the questions one asks!  This method {\em does not} address the question of
what regions of model space are allowed by the data, but rather what regions
provide a best fit under the assumption that the model is correct, for
{\em some} set of parameters.
The allowed region determined differs from that for the method outlined above
as it asks a different statistical question: not what models are allowed by
the data but what are the errors on the best-fit $\Delta m^2$ and
$\sin^2 2\theta$.

In addition, an approach for calculating allowed regions has recently been
advocated \cite{HatLan2} which uses non-standard definition of $\chi^2$.
In comparing their method to solar model Monte Carlos \cite{BU} the authors
define ``$\chi^2$'' in terms of the logarithm of the
``average probability'' rather than computing $V_{ab}$ for the
Bahcall \& Ulrich solar models directly and using equation (\ref{eqn:chi2def}).
The distribution of this ``$\chi^2$'' will not be chi-squared, and will not
take into account correlations in the rates in a well defined way.
To consistently use such a statistic, the correct distribution and confidence
levels to be associated with it would need to be calculated.

An alternative method, which is similar in spirit to the best fit approach,
is to calculate the 2D likelihood function
${\cal L}(\Delta m^2,\sin^2 2\theta)$, again under the assumption that the
variations in model predictions (and the experimental errors) are gaussian.
In this case the likelihood function is defined as\footnote{Note that the
authors of \cite{HatLan2} define a likelihood function as a sum of gaussians
and redefine $\nu$ above to give regions consistent with this approximate
likelihood function.  While the statistical meaning of this hybrid method
is not immediately clear, the final regions obtained are not much different
than provided by more conventional statistical treatments.}
\begin{equation}
  {\cal L} \propto {1\over\sqrt{\det V}} \exp\left[-{1\over 2}\chi^2\right]
\end{equation}
By use of Bayes'  theorem, the conditional probability for $\Delta m^2$ and
$\sin^2 2\theta$, given the experimental measurements, is proportional to the
likelihood function times the {\em a priori} probability distribution for the
MSW parameters \cite{Berger}
(for an alternative interpretation see \cite{Edwards}), which
is usually referred to as the posterior distribution.
If we assume, from scaling arguments, that logarithmic intervals in
$\Delta m^2$ and $\sin^2 2\theta$ are equally likely, before any experiment
is performed, then the posterior distribution is simply proportional to the
likelihood function ${\cal L}(\log\Delta m^2,\log\sin^2 2\theta)$.

To calculate the 95\% confidence regions for $\Delta m^2$ and $\sin^2 2\theta$
we follow the method used in assigning regions for gaussian distributions
(which ${\cal L}$ is not).  Let us define a region in parameter space
\begin{equation}
  A(\lambda) \equiv \left\{ (\log\Delta m^2,\log\sin^22\theta)
  \ |\ {\cal L}(\log\Delta m^2,\log\sin^22\theta) > \lambda \right\}.
\end{equation}
Then
\begin{equation}
  \Gamma(\lambda) \equiv \int_{A(\lambda)}\
  {\cal L}(\log\Delta m^2,\log\sin^22\theta)\
  d(\log\Delta m^2) d(\log\sin^22\theta)
\end{equation}
is a continuous, monotonic decreasing function of $\lambda$,
and the 95\% confidence region is given by $A(\lambda_{*})$, where
$\Gamma(\lambda_{*})=0.95\Gamma(0)$.
(We note that this method is somewhat arbitrary for multiply-peaked likelihood
functions such as ours, but it is nonetheless well defined.)
This confidence region is interpreted as the region that contains, with 95\%
probability, the true  values of $\Delta m^2$ and $\sin^2 2\theta$.
Although the interpretation of the region is different than that allowed by
the $\chi^2$ method, the two regions are encouragingly similar.
In the limit that the likelihood function were gaussian ($\chi^2$ is a
quadratic function of $\Delta m^2$ and $\sin^22\theta$ and $\det V$ is
constant) the regions would be ellipses as in \cite{HatLan2}.
Thus the departure from elliptical shape is an indication that the likelihood
function is not simply gaussian.

\vspace{\bigskipamount}

\noindent 6. {\bf Results}

Our principal result, the 95\% C.L.~allowed regions in MSW parameter space,
based only on statistical uncertainties in the present formulation of the
standard solar model, is shown in figure
\ref{fig:allowed} for both the
$\chi^2$ and
${\cal L}$ methods. The regions shown are obtained by requiring that
$\chi^2<9.49$ (4 dof), including both the rate and the day/night fits.
We also show the region obtained by requiring $\chi^2<\chi^2_{{\rm min}}+6.0$
(2 parameters) for comparison.

We see that there are two allowed regions, a large mixing angle (adiabatic)
region and a small mixing angle (non-adiabatic) region.
The small mixing angle region is favoured over the large mixing angle region,
though both are ``allowed'' at the 95\% confidence level.
Using the likelihood function we can ask what are the relative probabilities
of the large and small angle regions,
e.g.~we find $P(\sin^2 2\theta>0.1)\simeq 0.3P(\sin^2 2\theta<0.1)$
(see \cite{HatLan2} for a different way to ask this question).

One of the largest uncertainties in calculating the expected rates comes from
$S_{1,7}$, the nuclear cross section parameter for the reaction
${}^7$Be$(p,\gamma){}^8$B.
This uncertainty directly affects the flux of ${}^8$B neutrinos, and is due
to both experimental uncertainties and the difficulty of extrapolating the
experimental results to the low energies relevant in the solar interior.
There is a significant difference between the cross sections inferred from the
two experiments which have been performed at the lowest energies, indicating
some significant systematic error in this parameter.
The authors of ref \cite{BP} use a value for $S_{1,7}$ that is intermediate to
these two results, with errors which do not overlap the central values.
This value and its errors, which we have used in determining the correlation
matrix, {\em may} not properly reflect the uncertainty in $S_{1,7}$.
In order to explore the implications of a larger estimate for the error in
$S_{1,7}$, in figure \ref{fig:new-allowed} we present the allowed regions
assuming as the error for $S_{1,7}$ the difference between the two central
values of ref \cite{Kavanagh,Filippone}, which corresponds to a 21\%
uncertainty.  As is expected the allowed regions are correspondingly increased.

Recent refinements in solar models, i.e.~including heavy element diffusion,
have changed the predicted neutrinos fluxes from those of the Bahcall \&
Pinsonneault model.  As we have indicated, changes such as this, which includes
new physics rather than new numerical values for the input parameters and their
errors can have a large effect on the allowed regions.
The situation with respect to these new solar models is still not settled, and
the correlation matrix including parameters for the effects of heavy element
and helium diffusion are not yet available.
Nevertheless, to estimate the magnitude of the effect of such changes we have
use a hybrid procedure which uses neutrino fluxes including heavy element and
helium diffusion from \cite{Proffitt,Pins}, but our old correlation matrix.
Specifically we have artificially changed the fluxes of the
Bahcall \& Pinsonneault model by percentages equal to those shown in table
\ref{tab:newfluxes} but used the flux correlation matrix from table
\ref{tab:mc-correlation}.
While this method is not fully consistent, and the fluxes used are
preliminary, it should approximate the main effects of these changes and
illustrate the possible shift in the allowed regions that can be expected
for such models.

We display in figure \ref{fig:new-allowed} our result for the allowed range of
parameter space in this case.
The change in the allowed region reinforces our earlier remarks: the potential
change in the allowed regions due to such modifications of the solar model can
be larger than indicated by the inclusion of the usual solar model
uncertainties.
Thus it is prudent to realize that the presently ``allowed" regions are now
only suggestive.  Further solar model improvements could change their shape,
and position.

\vspace{\bigskipamount}

\noindent 7. {\bf MSW and Refined Confidence Limits: Future Work}

The statistical analysis we have provided here is straightforward, and
resolves various inconsistencies present in previous analyses.  As such,
it should provide a firm basis with which to analyse future results.
However, for the MSW solution, neither this approach, nor any other to date
actually properly accounts for all solar model uncertainties when determining
allowed ranges in parameter space.  As we have just indicated, statistical
solar model uncertainties do not incorporate possible systematic shifts in
fluxes due to new, non-exotic, physics, which could dramatically alter the
shape of allowed regions.
Beyond this, however, in the case of the MSW solution no set of neutrino flux
uncertainties can carry all of the relevant information on solar model
variations.
This is because the neutrino deficits which result from traversing the solar
interior themselves depend sensitively on the details of the solar density and
temperature.  Thus, simply calculating the initial neutrino fluxes over a wide
range of solar models, but propagating them using the density-temperature
relationship of the standard solar model is not fully consistent.
Stated in a language which can be compared with that given above, when
determining predicted rates in detectors $R_a$, we can no longer write
$R_a=r_{aj}\phi_j$ with $r_{aj}=r_{aj}(\Delta m^2,\sin^2 2\theta)$.
Rather, $r_{aj}$ now becomes also a function of $N_e(r)$,
$r_{aj}=r_{aj}(\Delta m^2,\sin^2 2\theta,N_e)$, so that the decomposition
which lead to equation (\ref{eqn:rcovariance}) can no longer be carried out.
Instead, for each value of $\Delta m^2$ and $\sin^2 2\theta$ one must carry out
the average over solar models directly, accounting for propagation in the sun
in each model, in order to determine the proper predicted rate covariance
matrix.
We do not know {\em a priori} how large an effect such a detailed accounting
will produce, although there is some reason to expect it will not change the
allowed region drastically.
Computationally this is far more daunting, and we will report on this analysis
in a future publication \cite{GKPW} where we will carry out such a procedure
for the newest set of solar models of Bahcall \& Pinnsonault, in which heavy
metal diffusion is accounted for, and for which the predicted neutrino fluxes
appear to be somewhat larger.

\vspace{\bigskipamount}

\noindent 8. {\bf Future Experiments}

At very large mixing angles we expect that the $\nu_e$ survival probability
will be roughly independent of energy so that the spectrum of neutrinos seen
would be unchanged but for the normalization. In the adiabatic region the
existence of a resonance implies a large suppression of the $\nu_e$ survival
probability while the converse is true in the non-adiabatic region
\cite{TheBook}.
Hence we expect that for $(\Delta m^2,\sin^22\theta)$ in the small mixing
angle allowed region, the lower energy $pp$ and ${}^7$Be neutrinos, which
have energies that correspond to the adiabatic regime, will be preferentially
depleted, while the higher energy ${}^8$B neutrinos have a higher survival
probability due to nonadiabatic level jumping.
Two neutrino experiments currently under construction, the Sudbury Neutrino
Observatory (SNO) and SuperKamiokande, may have the ability to detect the
distortion in the ${}^8$B neutrino energy spectrum for small angle MSW
solutions.  For most of the small angle region, SNO should be able to
discern the spectral distortion, while it is very unlikely that the minimal
shape distortion produced by MSW parameters in the large angle region could be
detected.

SNO of course can also measure the ratio of charged current events to neutral
current events, which provides an indicator for $\nu_e$ oscillations into
another active neutrino species.  A ratio significantly less than that expected
for the SSM (i.e.~the ratio of the electron neutrino charged current to neutral
currents cross sections \cite{BKN}) would be a strong indication of MSW mixing.
However, the neutral current events are signaled by the production of a free
neutron, and the background for this process can be problematic.

The improved statistics of SuperKamiokande, which expects to see on the order
of 8000 events per year \cite{SuperK}, can be used to examine more closely the
effects of $\nu_e$ regeneration through the earth (see section 4c).
In figure \ref{fig:superkam}, we plot the contours for several values of
(day-night)/(day+night), superimposed upon the allowed regions.
Recall that our model for calculating the above ratio automatically assumes
an average of the year and the entire night and uses a very crude model
of the earth.
An analysis using the methods of section 4c on a more realistic model of the
earth could be performed, however figure \ref{fig:superkam} serves to show
the potential for narrowing the allowed regions if a positive day/night
variation is seen.
If SuperKamiokande does find a signal for day/night variation then binning the
data vs $\cos\delta_{\rm sun}$ (c.f.~eq.~\ref{eqn:day-night} with
$\delta_{\rm sun}=(\theta_0+\pi)/2$) would provide more information than our
simple average.

An experiment sensitive to ${}^7$Be neutrinos can potentially discriminate
between the large and small mixing angle solutions in addition to confirming
the depletion of the ${}^7$Be flux.
The predicted rate for Borexino \cite{Borexino} as a fraction of the standard
solar model rate is \cite{GelKwoRos}
\begin{equation}
  R_{\rm Borexino} = 0.787 P(\nu_e\rightarrow\nu_e ; {}^7{\rm Be}) + 0.213
\end{equation}
where the $0.213$ is absent for oscillations into a sterile neutrino.
For the two currently allowed MSW parameter regions, the predicted rates in
Borexino are shown in figure \ref{fig:borexino}.
A Borexino rate of less than 0.3 would provide not only a striking
confirmation of the solar neutrino problem, but also indicate the small mixing
angle region of the MSW solution, whereas rates between 0.5 and 0.8 would point
to the large mixing angle solutions.
A detected rate of about 0.35 of the standard solar model would not allow
discrimination between the two solutions, but would nonetheless be further
evidence in support of new neutrino physics.
However note that after 1yr of running Borexino can at best measure the rate to
$\sim30\%$.

\vspace{\bigskipamount}

\noindent 9. {\bf Implications for particle physics models}

If future experiments confirm the deficit of electron neutrinos indicated by
the current data, we would have the first (indirect) evidence of physics
beyond the standard model: neutrino masses.
Further solar neutrino studies coupled with upcoming neutrino oscillation
experiments \cite{shortbase} are the current best hope of seeing neutrino
masses in the cosmologically interesting range $\sum m_i=3-30$eV.  (The region
of mass-mixing angle space of interest for oscillations which may explain the
deficit of atmospheric muon neutrinos can be probed by several proposed long
baseline oscillation experiments \cite{longbase}.)
The mass-mixing angle parameters implied by the allowed regions shown in
figure \ref{fig:allowed}, while not indicative of any particular particle
physics models for neutrino masses, are consistent with models which
incorporate a seesaw mechanism.
Many of these models can also accommodate the observed deficit of the ratio of
atmospheric $\nu_{\mu}/\nu_e$ \cite{mudef} and in some cases also allow for the
$\nu_\tau$ to be cosmological hot dark matter \cite{mdm} or provide
contributions to neutrinoless double $\beta$-decay at the level of
$\overline{m}\sim1$eV \cite{beta}.

There is a significant literature in the particle physics community on
constructing models which go beyond the Standard Model of Electroweak
interactions and many of these models have interesting implications for
neutrino properties.  Here we discuss some classes of models which are
currently popular and which relate directly to the solar neutrino problem.

Models in which the neutrino mixing angles are similar to the CKM angles in
the quark sector \cite{LanLuo} now appear to be disfavoured by the
data.
A class of models based on grand unification particle physics models and a
see-saw mechanism for neutrino masses \cite{gellmann} give masses
and mixings which can lie in the small angle region of figure \ref{fig:allowed}
\cite{Harvey,BabSha}.
In some cases these models can also incorporate solutions to the atmospheric
neutrino deficit, provide the hot dark matter component of currently popular
mixed dark matter models (on the order of 1-10 eV in neutrino mass),
and even accommodate a Majorana mass of 1-2 eV for neutrinoless double
$\beta$ decay \cite{moha1}.
In these models the masses of the light neutrinos we see are a combination of
the Dirac masses of the usual neutrinos plus new right handed neutrinos $\nu_R$
which additionally have Majorana masses.
The $\nu_R$ are placed in GUT gauge group multiplets along
with the quarks and leptons and get masses from the same Higgses.
This relates the Dirac mass matrices of the neutrinos in these models to those
of the quarks and leptons.
If further ``textures'' are assumed for the heavy Majorana mass matrix of the
$\nu_R$, one obtains predictions for the masses and mixings of the observed
light neutrinos, usually with one free (overall mass) scale and a small
number of group theory factors.
We note in passing that the presence of (powers of) these group theory factors
can significantly alter the naive see-saw predictions.

Other models exist \cite{Smirnov,PetSmi,KimLee,Lavoura}
which generate masses and mixing angles in the large angle allowed region of
\ref{fig:allowed}.
Such models can allow for simultaneous solution of the solar and atmospheric
neutrinos \cite{KimLee} or link solar neutrinos with double $\beta$-decay
experiments \cite{PetSmi} or both \cite{Bur}.

Some authors have considered a radiative mechanism for the generation of
neutrino masses, and found models which can accomodate two of the three
neutrino mass solutions (solar neutrinos, atmospheric neutrino deficit, dark
matter) \cite{zee}.

Further constraints on models for neutrino masses which invoke oscillations
into sterile neutrinos $\nu_s$ are obtained by considering big bang
nucleosynthesis \cite{BBN}.  The large angle region is excluded for
$\nu_e-\nu_s$ oscillations based on present observations of the primordial
${}^4$He abundance.
This also eliminates $\nu_{\mu}-\nu_s$ solutions to the atmospheric
$\nu_{\mu}$ deficit.
Arguments derived from supernova considerations can also be used to constrain
oscillations \cite{Qian,Sigl}.  For sterile neutrinos the region of mass-mixing
angle space restricted by these arguments, while of interest for sterile
neutrinos as dark matter candidates, is not relevant for solar neutrino
oscillations \cite{Sigl}.

\vspace{\bigskipamount}

\noindent 10. {\bf Conclusions}

In this paper we have presented an updated analysis of the implications of the
four currently operating solar neutrino experiments.  Our analysis incorporates
a straightforward and comprehensive treatment of the known theoretical
statistical uncertainties, which we have outlined in detail.
We have given a full account of our methods and assumptions so others can
compare with our work.

We find that the current solar neutrino experiments provide a useful constraint
on the masses and mixing angles of neutrino in models where neutrino mixing is
the resolution of the solar neutrino problem.  All the quantifiable errors in
established solar models are included in this constraint.  Both resonant (MSW)
and non-resonant (just-so) neutrino oscillation models are allowed by the data.
In considering the implications of these figures, it is important to note that
systematic uncertainties remaining in both the solar model calculations and
input parameters can have an effect on the properties of the allowed
regions, as shown in figure \ref{fig:new-allowed}.
Nevertheless, as solar models improve, the consistent statistical analysis we
have defined here will continue to gain in significance.

Future experiments have great promise for confirming the solar neutrino problem
and firmly establishing the need for neutrino based solutions.  In particular
we have examined the potential of SuperKamiokande, SNO and Borexino to provide
further constraints on the masses and mixing angles of neutrinos in such
models.  If the charged to neutral current ratio measured by SNO indicates the
probability of neutrino oscillations, the presence (absence) of spectral
distortion will further constrain the mixing parameters to the small (large)
angle regions.
(While SNO is not capable of distinguishing between large angle oscillations
into sterile neutrinos and solar model solutions, such oscillations have
already been ruled out by big bang nucleosynthesis as mentioned above
\cite{BBN}.)
Borexino also possesses the ability to distinguish between small and large
angle
MSW regions to some extent.  If the large angle solution turns out to be
favoured, then SuperKamiokande should provide a sensitive probe of the allowed
mass and mixing through the measurement of the day/night effect.

\vspace{\bigskipamount}

\noindent {\bf Acknowledgements}

We would like to thank J. Bahcall for kindly providing us with the total
fluxes from the $1{,}000$ solar models of \cite{BU}, and for helpful
discussions.
We also thank M. Pinnsonault and N. Hata for useful conversations related to
their work. Work supported in part by the NSF and DOE.
M.W.~acknowledges the support of an SSC fellowship from the TNRLC.
L.M.K.~and M.W.~thank the Institute for Nuclear Theory at the University
of Washington for its hospitality and the Department of Energy for partial
support during the completion of this work. E.G. is supported by the
DOE at the University of Chicago and by the DOE and by NASA through
grant NAGW2381 at Fermilab.

\clearpage

\begin{table}
\begin{center}
\begin{tabular}{|c|c|c|c|}
\hline
&& \\
$R/R_{\oplus}$ & $\rho$ & $\langle \sqrt{2}G_F N_e \rangle$ \\
&& \\ \hline
0.0000-0.1910 & 12.858 & 4.87 \\
0.1910-0.5471 & 11.024 & 4.18 \\
0.5471-0.8948 &  4.964 & 1.88 \\
0.8948-0.9341 &  3.923 & 1.49 \\
0.9341-1.0000 &  2.292 & 0.87 \\ \hline
\end{tabular}
\end{center}
\caption{The model Earth that we used in calculating the regeneration effect.
Densities in column 2 are given in $g/{\rm cm}^3$ and the final column is
in $10^{-7}{\rm eV}^2$/MeV, assuming $n_p=n_n$ for the Earth interior.}
\label{tab:earthmodel}
\end{table}

\begin{table}
\begin{center}
\begin{tabular}{|l|c|}
\hline
Experiment & Rate  \\ \hline
Homestake  & $0.31 \pm 0.03$ \\
Kamiokande & $0.51 \pm 0.07$ \\
Gallium    & $0.59 \pm 0.08$ \\ \hline
\end{tabular}
\caption{The experimental rates, normalized to the standard solar model
predictions, used in the fits.  The rates for SAGE and GALLEX have been
combined and cross section uncertainties for Homestake and Gallium have
been added, in quadrature, to the experimental errors.}
\end{center}
\label{tab:exptrates}
\end{table}

\begin{table}
\begin{center}
\begin{tabular}{|c|ccccccccccc|} \hline
 & H & K & Ga & pp & pep & hep & ${}^7$Be & ${}^8$B & ${}^{13}$N & ${}^{15}$O
 & ${}^{17}$F \\ \hline
         H&100& 99& 97&-&-&-&-&-&-&-&- \\
         K& 99&100& 95&-&-&-&-&-&-&-&- \\
        Ga& 97& 95&100&-&-&-&-&-&-&-&- \\
        pp& - & - & - &100& 77& 16&-91&-72&-88&-88&-88 \\
       pep& - & - & - & 77&100& 28&-69&-50&-73&-71&-72 \\
       hep& - & - & - & 16& 28&100&  5&-15&-46&-45&-45 \\
  ${}^7$Be& - & - & - &-91&-69&  5&100& 74& 80& 80& 80 \\
   ${}^8$B& - & - & - &-72&-50&-15& 74&100& 73& 73& 73 \\
${}^{13}$N& - & - & - &-88&-73&-46& 80& 73&100& 99& 99 \\
${}^{15}$O& - & - & - &-88&-71&-45& 80& 73& 99&100& 99 \\
${}^{17}$F& - & - & - &-88&-72&-45& 80& 73& 99& 99&100 \\ \hline
\end{tabular}
\end{center}
\caption{The experiment and flux correlations ($\times 100$) computed using
the $1{,}000$ solar models of Bahcall \& Ulrich.}
\label{tab:bu-correlation}
\end{table}

\begin{table}
\begin{center}
\begin{tabular}{|c|ccccccccccc|} \hline
 & H & K & Ga & pp & pep & hep & ${}^7$Be & ${}^8$B & ${}^{13}$N & ${}^{15}$O
 & ${}^{17}$F \\ \hline
         H&100& 99& 95& - & - & - & - & - & - & - & -  \\
         K& 99&100& 92& - & - & - & - & - & - & - & -  \\
        Ga& 95& 92&100& - & - & - & - & - & - & - & -  \\
        pp& - & - & - &100& 77&  8&-90&-74&-88&-88&-88 \\
       pep& - & - & - & 77&100& 12&-70&-53&-71&-69&-71 \\
       hep& - & - & - &  8& 12&100&  2& -6&-21&-20&-21 \\
  ${}^7$Be& - & - & - &-90&-70&  2&100& 75& 77& 77& 80 \\
   ${}^8$B& - & - & - &-74&-53& -6& 75&100& 73& 73& 75 \\
${}^{13}$N& - & - & - &-88&-71&-21& 77& 73&100&100& 97 \\
${}^{15}$O& - & - & - &-88&-69&-20& 77& 73&100&100& 96 \\
${}^{17}$F& - & - & - &-88&-71&-21& 80& 75& 97& 96&100 \\ \hline
\end{tabular}
\end{center}
\caption{The experiment and flux correlations ($\times 100$) computed using
the power-law Monte Carlo approach.}
\label{tab:mc-correlation}
\end{table}

\begin{table}
\begin{center}
\begin{tabular}{|c|c|c|c|}
\hline
Flux       & \% Change \\ \hline
$pp$       &  1$\downarrow$ \\
pep        &  2$\downarrow$ \\
hep        &  2$\downarrow$ \\
${}^7$Be   &  5$\uparrow$ \\
${}^8$B    & 14$\uparrow$ \\
${}^{13}$N &  4$\uparrow$ \\
${}^{15}$O & 24$\uparrow$ \\ \hline
\end{tabular}
\end{center}
\caption{The percentage change in the fluxes for each neutrino species in
the Proffitt solar model arising from including heavy element diffusion.}
\label{tab:newfluxes}
\end{table}

\clearpage

\begin{figure}[htb]
\begin{center}
\leavevmode
\end{center}
\caption{The survival probability $P(\nu_e\rightarrow\nu_e)$ including
the Earth effect for mixing angle $\sin2\theta=0.4$.
The dotted line is the MSW probability without the Earth effect, the
solid line is the probability for a neutrino passing through the center
of the Earth and the dot-dashed line shows the probability averaged over
the night/year.}
\label{fig:regeneration}
\end{figure}

\begin{figure}[htb]
\begin{center}
\leavevmode
\end{center}
\caption{Region of MSW mass-mixing angle space allowed at the 95\% confidence
level for the combined Homestake-Kamiokande-Gallium data including solar model
uncertainties from the $\chi^2$ analysis (solid).
Also plotted is the 95\% confidence region from the likelihood function
analysis (dotted) and the region obtained by requiring
$\chi^2<\chi^2_{{\rm min}}+\nu$ (dashed).}
\label{fig:allowed}
\end{figure}

\begin{figure}[htb]
\begin{center}
\leavevmode
\end{center}
\caption{Likelihood function
${\cal L}(\log\Delta m^2,\log\sin^2 2\theta)$ for the combined
Homestake-Kamiokande-Gallium data.}
\label{fig:likelihood}
\end{figure}

\begin{figure}[htb]
\begin{center}
\leavevmode
\end{center}
\caption{Region of MSW mass-mixing angle space allowed at the 95\% confidence
level in the Bahcall \& Pinsonneault standard model for the combined
Homestake-Kamiokande-Gallium data including normal solar model uncertainties
for $\chi^2<9.49$ (solid).
Also plotted is the 95\% confidence region increasing the error on $S_{1,7}$
to 21\% (dotted) and the region obtained for fluxes approximating the effects
of metal diffusion (dashed, see text).}
\label{fig:new-allowed}
\end{figure}

\begin{figure}[htb]
\begin{center}
\leavevmode
\end{center}
\caption{The allowed region from the fit to the experimental rates, plus the
contours of (day-night)/(day+night) rate.  The contours are 1\% (solid),
5\% (dotted), 10\% (dashed) and 15\% (long-dashed).  An average over the
night and the year has been assumed for these contours, and the model of
the interior of the earth used was very simplistic.}
\label{fig:superkam}
\end{figure}

\begin{figure}[htb]
\begin{center}
\leavevmode
\end{center}
\caption{Predictions for the Borexino event rate as a fraction of the
standard solar model value.  The histograms show relative frequencies
of predicted event rates in the large (dashed) and small (solid) angle
regions.}
\label{fig:borexino}
\end{figure}


\clearpage

\begin{thebibliography}{99}

\frenchspacing

\bibitem{White} M. White, L.M. Krauss and E. Gates, Phys. Rev. Lett. {\bf 70}
(1993) 375

\bibitem{BahcallBethe} H.A. Bethe and J.N. Bahcall, Phys. Rev. {\bf D47} (1993)
1298

\bibitem{Bah} J.N. Bahcall, Phys. Rev. {\bf D44} (1991) 1644;
H.A. Bethe and J.N. Bahcall, Phys. Rev. {\bf D44} (1991) 2962;
J.N. Bahcall and H.A. Bethe, Phys. Rev. Lett {\bf 65} (1990) 2233;
J.N. Bahcall and W.C. Haxton, Phys. Rev. {\bf D40} (1989) 931

\bibitem{BKL} S.A. Bludman, D.C. Kennedy and P.G. Langacker,
Phys. Rev. {\bf D45} (1992) 1810;
Nucl. Phys. {\bf B374} (1992) 373;
Nucl. Phys. {\bf B373} (1992) 498;
Nucl. Phys. (Supp.) {\bf B31} (1993) 156

\bibitem{ShiSchBah} X. Shi, D.N. Schramm and J.N. Bahcall,
Phys. Rev. Lett. {\bf 69} (1992) 717;
X. Shi and D.N. Schramm, Phys. Lett. {\bf B283} (1992) 305;
X. Shi, D.N. Schramm and D.S.P. Dearborn, FNAL-Pub-94/122-A (1994)

\bibitem{KraPetMSW} P.I. Krastev and S.T. Petcov,
Phys. Lett. {\bf B299} (1993) 99

\bibitem{KwoRos} W. Kwong, S.P. Rosen, UTAPHY-HEP-9

\bibitem{BHKL}  S.A. Bludman, N. Hata, D.C. Kennedy, and P.G. Langacker,
Phys. Rev {\bf D47} (1993) 2220

\bibitem{MSW} S.P. Mikheyev, A. Yu. Smirnov, Sov. J. Nucl. Phys.
{\bf 42} (1985) 913;
Nuo. Cim. {\bf 9C} (1986) 17;
L. Wolfenstein,  Phys. Rev. {\bf D17} (1987) 2369;
{\bf D20} (1989) 2634

\bibitem{HatLan2} N. Hata and P. Langacker, U. Penn preprint UPR-0592T (1993)

\bibitem{BU} J.N. Bahcall and R.K. Ulrich, Rev. Mod. Phys. {\bf 60} (1988) 297

\bibitem{TheBook} J. N. Bahcall, ``Neutrino Astrophysics", Cambridge
University Press, (1989)

\bibitem{statbook} J. Mathews and R.L. Walker, ``Methods of Mathematical
Physics'', 2nd ed., Addison-Wesley (1970);
T.J. Loredo, ``Statistical Challenges in Modern Astronomy'',
E.D. Feigelson and G.J. Babu (eds.), Springer-Verlag, New York (1992).

\bibitem{KGW} L.M. Krauss, E. Gates, and  M. White, Phys. Lett. B {\bf 299}
(1993) 94

\bibitem{BP} J.N. Bahcall and M. H. Pinsonneault, Rev. Mod. Phys. {\bf 64}
(1992) 885

\bibitem{vacuum} V. Gribov and B. Pontecorvo, Phys. Lett. {\bf B28} (1969) 493;
S.M. Bilenky and B. Pontecorvo, Phys. Rep. {\bf 41} (1978) 225;
S.M. Bilenky and S.T. Petcov, Rev. Mod. Phys. {\bf 59} (1987) 671;

\bibitem{KPvac} P. I. Krastev and S. T. Petcov,
Phys. Rev. Lett. {\bf 72} (1994) 1960

\bibitem{Hata} N. Hata, U. Penn preprint, UPR-0605T (1994)

\bibitem{Homestake} R. Davis, Jr., in ``Proceedings of the Seventh Workshop
on Grand Unification",
Toyama, 1986, p.237, ed J. Arafune, World Scientific, (1986);
in ``Frontiers of Neutrino Astrophysics'', Tokyo, 1993, p.47,
ed. Y. Suzuki and K. Nakamura, Universal Academy Press (1993)

\bibitem{Kamiokande1} K. Hirata et al., Phys. Rev. Lett. {\bf 63} (1989) 16;
Phys. Rev. Lett. {\bf 65} (1990) 1297;
Phys. Rev. Lett. {\bf 65} (1990) 1301;
Phys. Rev. Lett. {\bf 66} (1991) 9;
A. Suzuki, KEK preprint 93-96 (1993)

\bibitem{Sage} A.I. Abazov et al., Phys. Rev. Lett. {\bf 67} (1991) 3332;
Nucl. Phys. (Supp.) {\bf B19} (1991) 84

\bibitem{Gallex} P. Anselmann et al., Phys. Lett. B {\bf 285}
(1992) 376; {\bf 285} (1992) 390; {\bf B314} (1993) 445

\bibitem{KP} P. I. Krastev and S. T. Petcov, Phys. Lett. {\bf B207} (1988) 64

\bibitem{probs} W.C. Haxton, Phys. Rev. {\bf D35} (1987) 2352;
S.J. Parke, Phys. Rev. Lett. {\bf 57} (1986) 1275;
P. Pizzochero, Phys. Rev. {\bf D36} (1987) 2293;
T.K. Kuo and J. Pantaleone, Rev. Mod. Phys. {\bf 61} (1989) 937

\bibitem{earth} E.D. Carlson, Phys. Rev. {\bf D34} (1986) 1454;
J. Boucher, M.  Cribier, W. Hampel, J. Rich, M. Spiro, and D. Vignaud,
Z. Phys. {\bf C32} (1986) 499

\bibitem{HatLan1} N. Hata and P. Langacker, Phys. Rev. {\bf D48} (1993) 2937

\bibitem{BalWen}A.J. Baltz and J. Weneser, Phys. Rev {\bf D35} (1987) 528;
{\bf D37} (1988) 3364

\bibitem{earthmodel} A.M. Dziewonski,, A.L. Hales and E.R. Lapwood, Phys. Earth
Plan. Int. {\bf 10} (1975) 12; F.D. Stacey, {\it Physics of the Earth} 2nd Ed.
(Wiley, New York, 1985)

\bibitem{Gates} E. Gates, L.M. Krauss and M. White, Phys. Rev. {\bf D46}
(1992) 1263

\bibitem{Hirata} K. Hirata, et al., Phys. Rev. {\bf D44} (1991) 2241;
M. Nakahata, Ph.D Thesis, ICEPP preprint UT-ICEPP-88-1

\bibitem{statart} see, for example:
E.L. Lehmann, ``Testing Statistical Hypotheses'', Wiley-Interscience (1986);
A.C.S. Readhead and C.R. Lawrence, Ann. Rev. Astron. \& Astrophys
{\bf 30} (1992) 653;
A.C.S. Readhead, et al., Ap.J. {\bf 346} (1989) 566, section VIII;
``Review of Particle Properties'', Phys. Rev. {\bf D45} (1992) Part II;
C. Howson and P. Urbach, Nature {\bf 350} (1991) 371;
J. Skilling, Nature {\bf 353} (1991) 707;
A.W.F. Edwards, Nature {\bf 352} (1991) 386;

\bibitem{Berger} J.O. Berger, ``Statistical decision theory and Bayesian
analysis'', 2nd ed., Springer-Verlag, (1985)

\bibitem{Edwards} A.W.F. Edwards, ``Likelihood'', Cambridge UP (1984)

\bibitem{Kavanagh} R. W. Kavanagh {\it et al.}, Bull. Am. Phys. Soc. {\bf 14}
(1969) 1209

\bibitem{Filippone} B. W. Filippone {\it et al.}, Phys. Rev. {\bf C28} (1983)
2222

\bibitem{GKPW} E. Gates, L.M. Krauss, M. Pinsonneault and M. White,
in preparation

\bibitem{Proffitt} C. Proffitt, talk given at ``Solar Modelling Workshop'',
March 1994, (INT, Seattle).

\bibitem{Pins} M. Pinsonneault, private communication.

\bibitem{SNO} G.T. Ewan et al., in Frontiers of Neutrino Astrophysics,
Y. Suzuki and K. Nakamura (Eds.), Universal Academy Press, Inc. Tokyo,
Japan (1993) 147

\bibitem{BKN} J.N. Bahcall, K. Kubodera, and S. Nozawa, Phys. Rev. {\bf D38}
(1988) 1030

\bibitem{SuperK} M. Takita in Frontiers of Neutrino Astrophysics,
Y. Suzuki and K. Nakamura (Eds.), Universal Academy Press, Inc. Tokyo,
Japan (1993) 135

\bibitem{Borexino} R.S. Raghavan, et al., Phys. Rev. {\bf D44} (1991) 3786

\bibitem{GelKwoRos} J.M. Gelb, W. Kwong and S.P. Rosen,
Phys. Rev. Lett. {\bf 69} (1992) 1864;
W. Kwong and S.P. Rosen, Phys. Rev. Lett. {\bf 68} (1992) 748

\bibitem{shortbase} N. Armenise et al., CERN-SPSC/90-42 (1990) (CHORUS);
K. Kodama et al. (E803), FNAL proposal (October 1993);
P. Astier et al., CERN-SPSLC/92-21 (1992) (NOMAD)

\bibitem{longbase} S. Parke,
``Overview of Accelerator Long Baseline Neutrino Oscillation
Experiments'' in {\em Perspectives in Neutrinos, Atomic
Physics and Gravitation}, proceedings of 1993 Moriond Meeting, J.
Tran Thanh Van et al, eds, Editions Frontieres, Gif-sur-Yvette, 1993.

\bibitem{mudef} K.S. Hirata, et al., Phys. Lett. {\bf B280} (1992) 146;
R.  Becker-Szendy et al., Phys. Rev. {\bf D46} (1992) 3720;
P.J. Litchfield in International Europhysics Conference on High Energy Physics,
Marseille, France (1993) (unpublished)

\bibitem{mdm} Q. Shafi and F. Stecker, Phys. Rev. Lett. {\bf 53} (1984) 1292;
D. Tommasini, preprint FTUV/94--1

\bibitem{beta} A. Piepke in International Europhysics Conference on High Energy
Physics, Marseille, France (1993) (unpublished);
E. Garcia in TAUP 93 Workshop, Gran Sasso, Italy (1993) (unpublished)

\bibitem{LanLuo} P. Langacker \& M. Luo, Phys. Rev. {\bf D44} (1991) 817;
P. Langacker et al., Nucl. Phys. {\bf B282} (1987) 589

\bibitem{gellmann} M. Gell-Mann, P. Ramond, and R. Slansky, in Supergravity,
ed. F. van Nieuwenhuizen and D. Freedman (Norht-Holland, Amsterdam) (1979) 315

\bibitem{Harvey} J. Harvey, D. Reiss, and P. Ramond,
Phys. Lett. {\bf B92} (1980) 309;
Nucl. Phys. {\bf B199} (1982) 223

\bibitem{BabSha} K.S. Babu \& Q. Shafi, Phys. Rev. {\bf D47} (1993) 5004;
Phys. Lett. {\bf B194} (1992) 235;
S. Dimopoulos, L.J. Hall \& S. Raby, Phys.  Rev. {\bf D47} (1993) 3697;
K.S. Babu \& R.N. Mohapatra, Phys. Rev. Lett. {\bf 70} (1993) 2845;
G. Anderson et al., Phys. Rev. {\bf D49} (1994) 3660;
C.H. Albright \& S. Nandi, Fermilab-PUB-94/061-T;
E. Papageorgiu, preprint;D. Lee and R. N. Mohapatra, UMD-PP-94-95

\bibitem{moha1} D.O.Caldwell and R.N. Mohapatra,
Phys. Rev. {\bf D48} (1993) 3259;
D.O.Caldwell and R.N. Mohapatra, UCSB-HEP-94-03/UMD-PP-94-90 (1994)

\bibitem{Smirnov} A. Smirnov, Phys. Rev. {\bf D48} (1993) 3264

\bibitem{PetSmi} S.T. Petcov and A.Yu Smirnov,
Phys. Lett. {\bf B322} (1994) 109

\bibitem{KimLee} C.W. Kim \& J.A. Lee, JHU-TIPAC-930023

\bibitem{Lavoura} L. Lavoura, Phys. Rev. {\bf D48} (1993) 5440

\bibitem{Bur} C.P. Burgess, Phys. Rev. {\bf D48} (1993) 4326

\bibitem{specshape} H.A. Bethe, Phys. Rev. Lett. {\bf 56} (1986) 1305;
S.P. Rosen and J.M. Gelb, Phys. Rev. {\bf D34} (1986) 969;
E.W. Kolb, M.S. Turner and T.P. Walker, Phys. Lett. {\bf B175} (1986) 478

\bibitem{zee} A.Y. Smirnov and Z. Tau, IC preprint 94/42 (1994),
and references therein

\bibitem{BBN} X. Shi, D.N. Schramm, and B.D. Fields, Phys. Rev. {\bf D48}
(1993) 2563

\bibitem{Qian} Y.-Z Qian et al., Phys. Rev. Lett. {\bf 71} (1993) 1965

\bibitem{Sigl} X. Shi and G. Sigl, Phys. Lett. {\bf B323} (1994) 360

\nonfrenchspacing

\end{thebibliography}
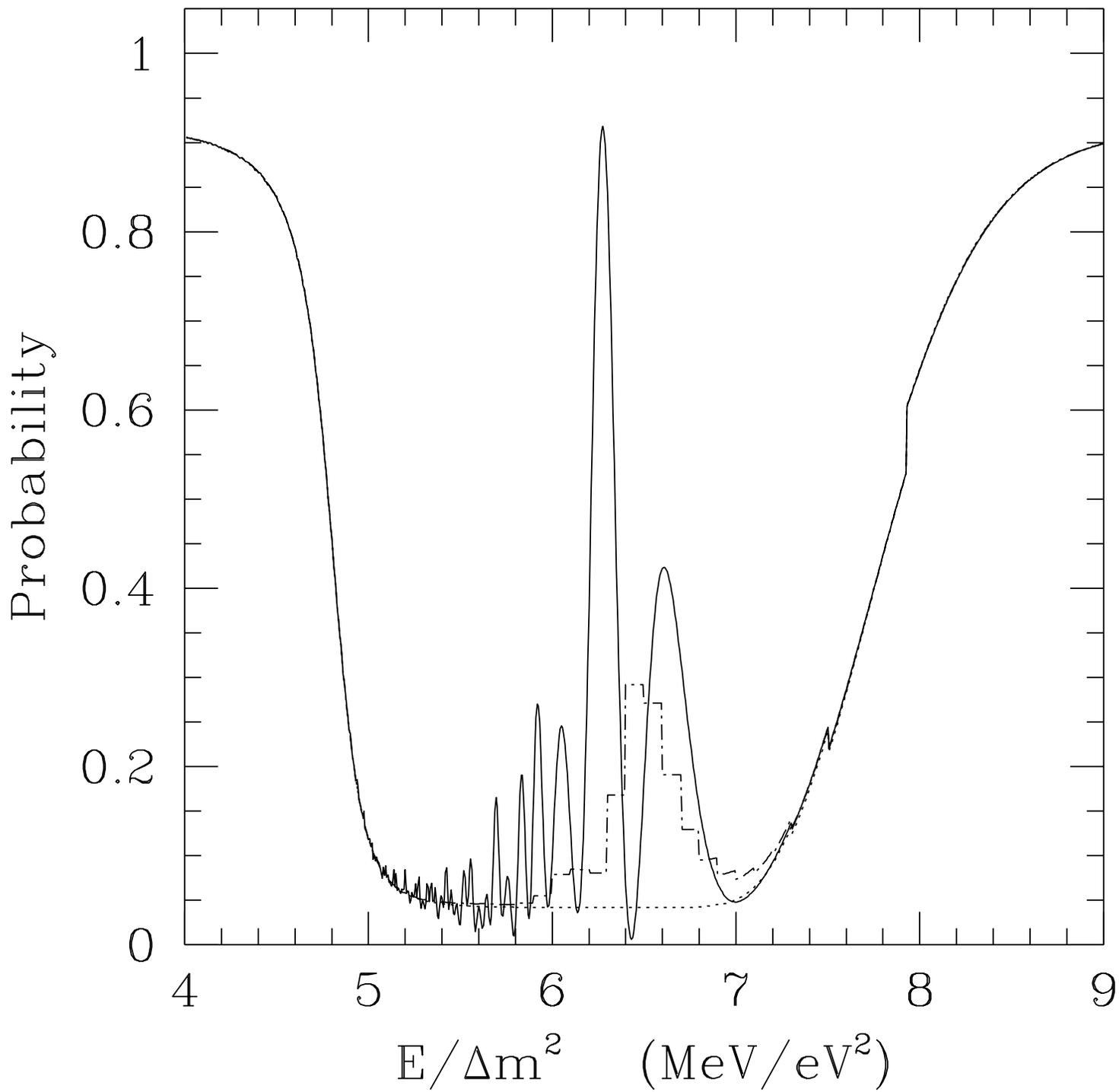
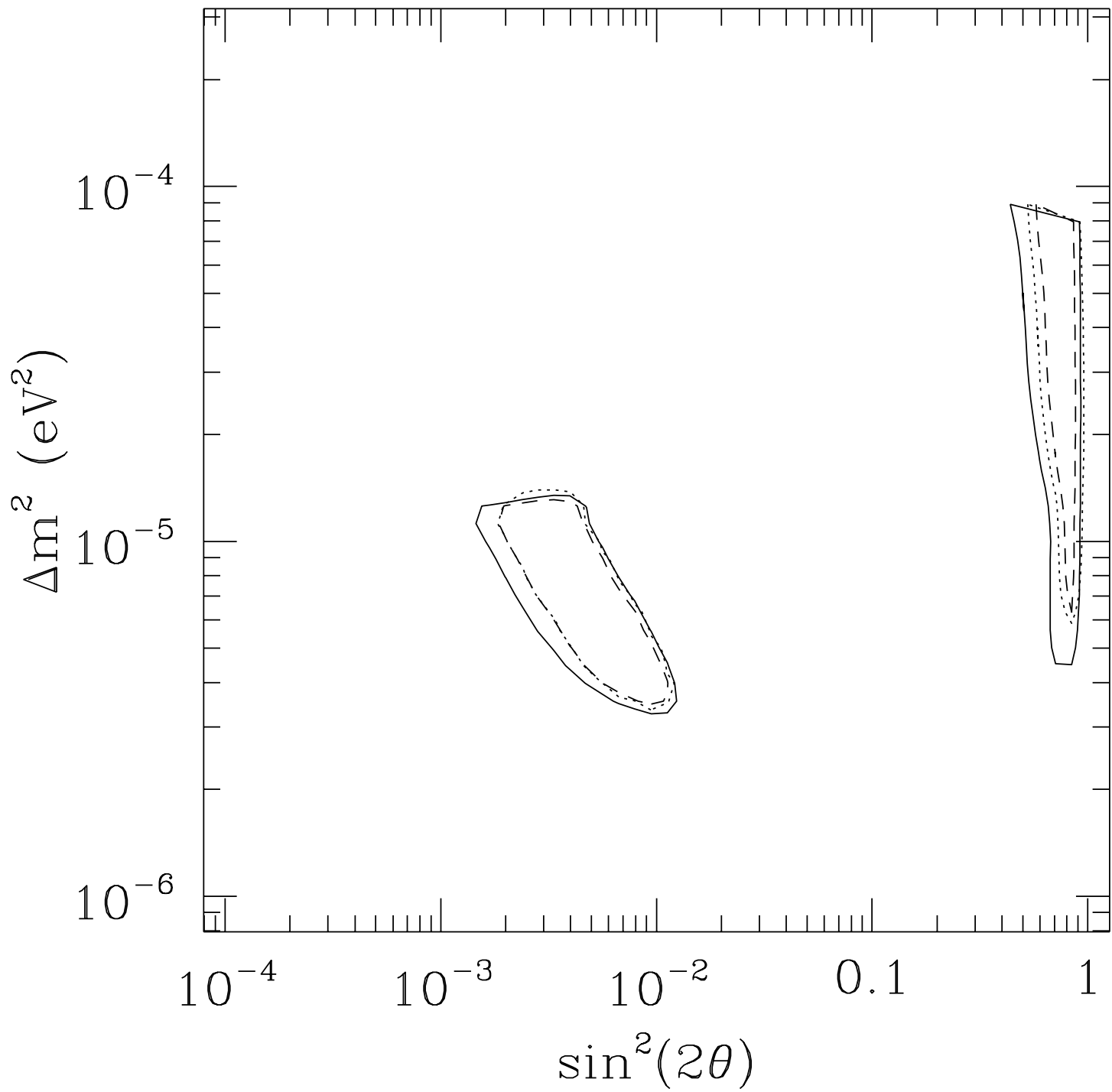
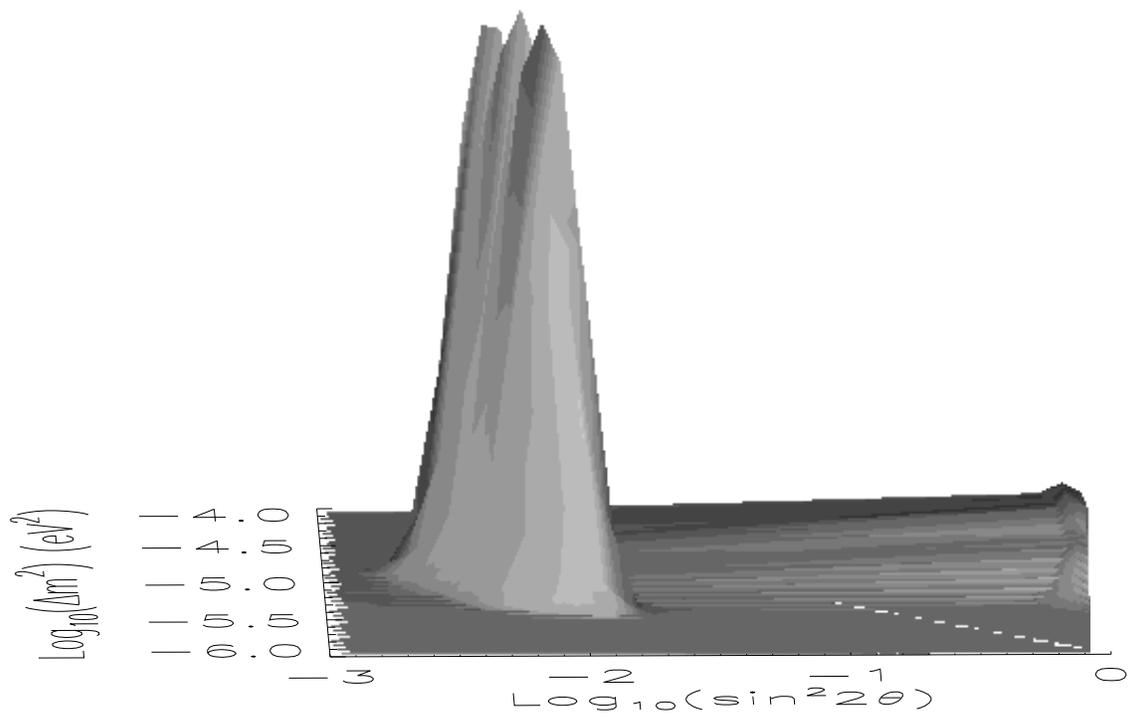
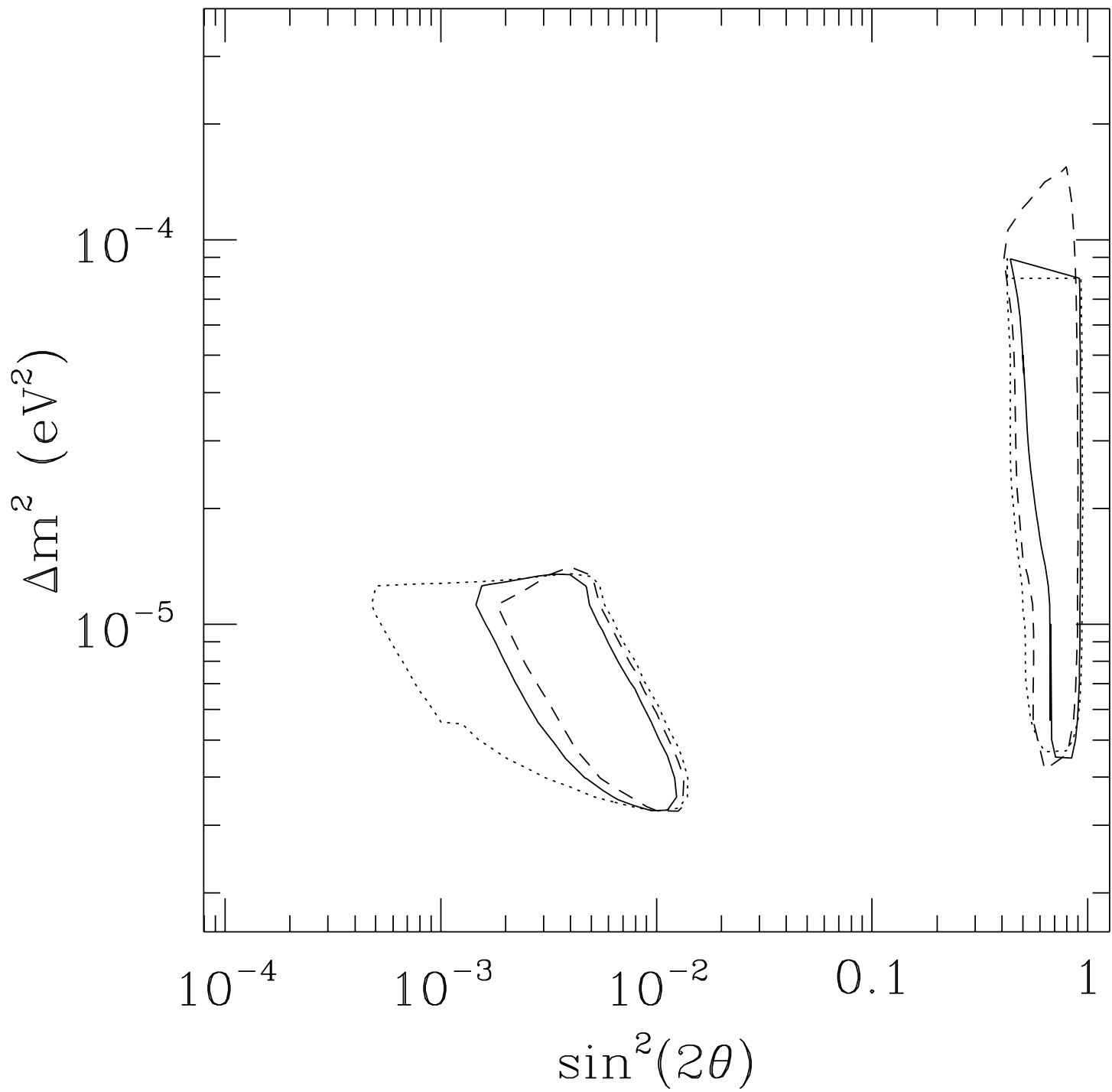
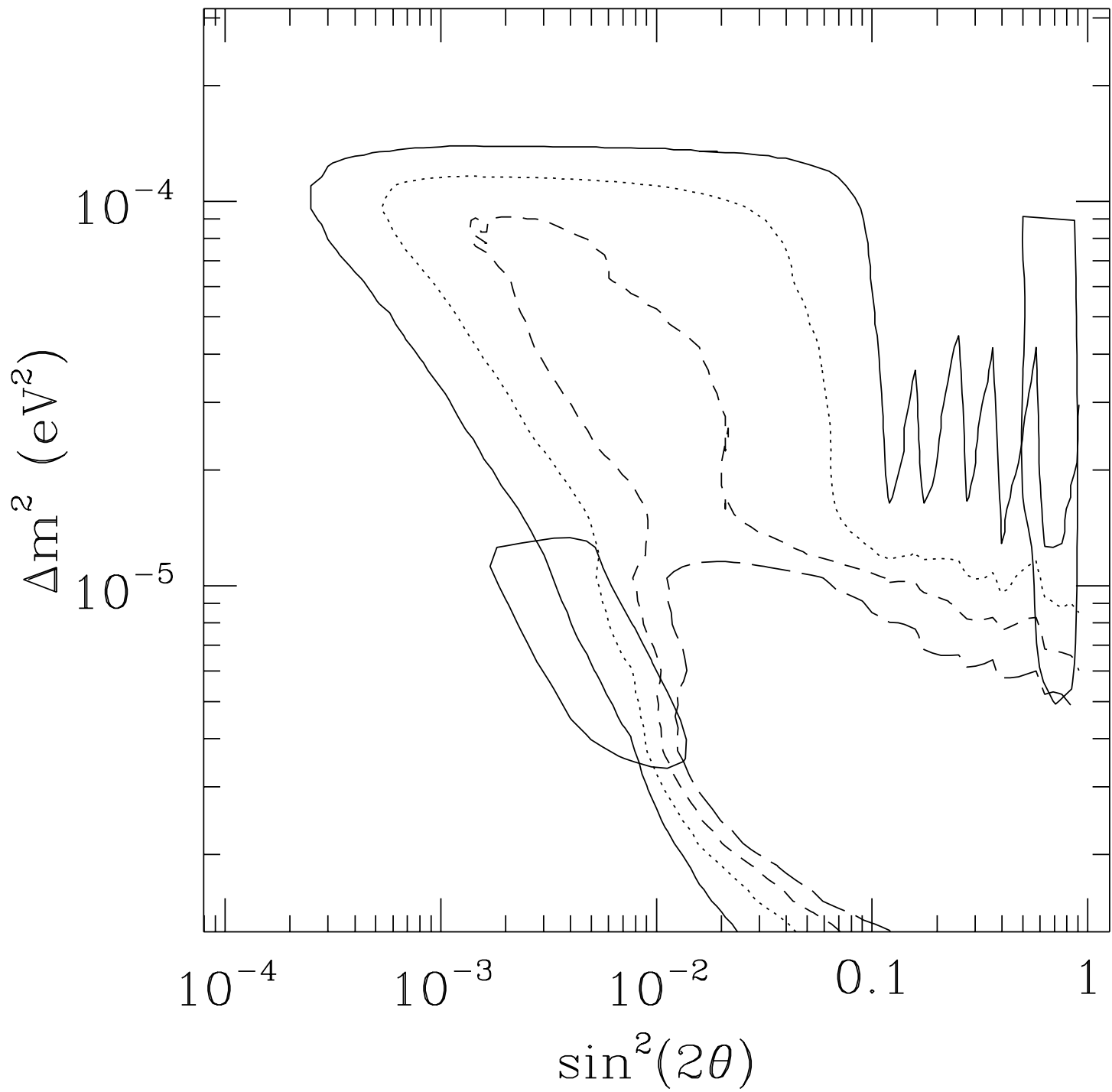
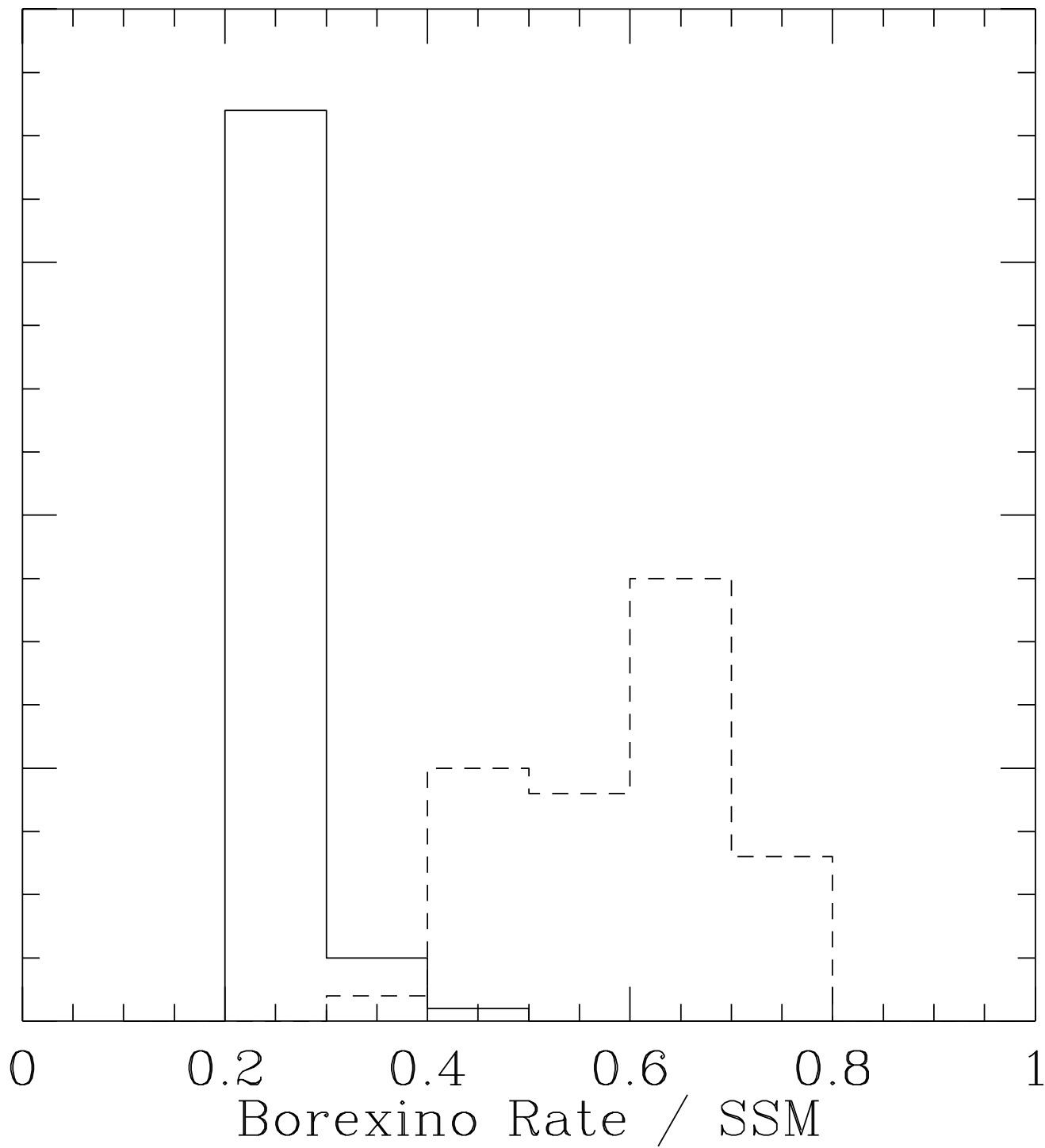
\end{document}